\documentclass{elsart1p}

\usepackage{graphics}
\usepackage{graphicx}
\usepackage{epsfig}

\usepackage{amssymb}
\begin{document}
\begin{frontmatter}

\title{Antikaon Interactions with Nucleons and Nuclei}
\author{Wolfram Weise}\footnote{Supported in part by BMBF, GSI and by the DFG cluster of excellence Origin and Structure of the Universe.}
\address{Physik-Department, Technische Universit\"at M\"unchen, D-85747 Garching, Germany}

\begin{abstract}
This report summarizes our understanding of $\bar{K}$-nucleon interactions and reviews the present theoretical situation in the quest for quasibound antikaon-nuclear systems.
\end{abstract}

\end{frontmatter}

\section{Introduction}
\label{Intro}
The investigation of antikaon (strangeness $S = -1$) interactions with nucleons and nuclei has a long and interesting history based on early observations that the isospin $I=0$ s-wave $\bar{K}N$ is quite strongly attractive around and below $\bar{K}N$ threshold, unlike the weakly repulsive kaon-nucleon ($S = +1$) interaction. The frame for this is chiral SU(3) effective field theory, the low-energy realization of QCD with strange quarks. This theory uniquely identifies the Tomozawa-Weinberg (TW) terms as the driving sources of the low-energy $\bar{K}N$ interaction. For example, the TW interaction Hamiltonian  in the $K^-p\rightarrow K^-p$ elasic channel at zero three-momentum is
\begin{center}
$\delta H = -(i/2f^2)\int d^3x\,\Psi_p^\dagger(x)\, K^+(x)\,\partial_t K^-(x)\, \Psi_p(x)~,$
\end{center}
where $K(x)$ and $\Psi_p(x)$ denote the kaon and proton fields. The $K^-n\rightarrow K^-n$ interaction is half as strong. The coupling strength to this order is solely determined by the pseudoscalar decay constant, $f \simeq 0.1$ GeV. Note that this interaction is proportional to the kaon energy. It vanishes in the chiral limit (i.e. for vanishing quark masses) and at zero kaon energy $(\omega = 0)$, as it should for a Goldstone boson. Explicit chiral symmetry breaking by the strange quark mass gives the charged kaon its observed mass, $m_K\simeq 494$ MeV, the energy scale around which threshold $\bar{K}N$ physics takes place. At this energy scale the resulting isospin-zero s-wave $\bar{K}N$ force is indeed quite strongly attractive, resulting in a $K^-$-nuclear potential of order -100 MeV at the center of heavy nuclei. While higher order terms in the low-energy expansion of the chiral effective Lagrangian are not negligible, the TW term nonetheless dominates the $\bar{K}N$ s-wave. 

Historically, ideas  concerning $K^-$ condensation in neutron star matter \cite{KN86} started from this observation. Calculations of the antikaon spectrum in dense nuclear matter \cite{WKW96} display the expected attractive shift of the antikaon mass. Such early calculations suggested that the in-medium $\bar{K}$ mass shift, or binding energy, might reach 20 \% of the vacuum kaon mass at normal nuclear matter density, while at the same time the ${\bar K}N \rightarrow \pi\Sigma$ decay width drops once the phase space for this process closes. Exploratory studies for neutron star matter were performed in the same framework \cite{WRW97}. Such calculations suggested typical densities for (anti)kaon condensation, at which the in-medium effective $K^-$ mass meets the electron chemical potential, around four times the density of normal nuclear matter.  

Of course, the model dependence of such extrapolations to dense matter remained an issue ever since. The role of the $\bar{K}NN\rightarrow YN$ absorption channels, producing hyperons rather than antikaonic modes in the ground state of the dense medium, needs still to be clarified quantitatively. Refined calculations were performed, e.g. in Ref. \cite{LK08}, implementing partial self-consistency for the in-medium $\bar{K}$ propagator. The data base and phenomenology of kaonic atoms was systematically investigated to set empirical constraints on the attractive $K^-$-nuclear forces \cite{FG07}. 
While the qualitative consensus about the attractive nature of antikaon-nuclear forces is without doubt, quantitative predictions about the far-subthreshold properties of this interaction are still unavoidably vague and will presumably remain so until fully conclusive experimental data can be used to set more stringent constraints. The present report attempts to give a state-of-the-art picture of recent theoretical developments. Reports on the experimental status are found elsewhere in these proceedings \cite{CP09}.

\section{Low-energy antikaon-nucleon interactions}

The chiral  SU(3) meson-baryon effective Lagrangian is the appropriate starting point for a systematic construction of leading low-energy $\bar{K}N$ interaction terms. However, chiral perturbation theory is not applicable in the sector with baryon number $B = 1$ and strangeness $S = -1$. The reason is the formation of the $\Lambda(1405)$ at less than 30 MeV below $K^-p$ threshold. The strong decay of the $\Lambda(1405)$ into $\pi\Sigma$ immediately implies a coupled-channels problem that must be solved to all orders. This was recognized already long ago by R. Dalitz and collaborators \cite{Da67}. They were the first to suggest that the $\Lambda(1405)$ is not a simple three-quark baryon but rather a $\bar{K}N$ quasibound state embedded in the $\pi\Sigma$ continuum. The contemporary framework to approach this physics is chiral SU(3) dynamics \cite{KSW95, OR98, LK02}. It combines the non-perturbative coupled-channels method with input from the chiral SU(3) Lagrangian, 
\begin{center}${\cal L}_{eff} = {\cal L}_{mesons}(\Phi) + {\cal L}_B(\Phi,\Psi_B)$,\end{center} that involves the pseudoscalar meson octet fields $(\Phi)$ interacting with the baryon octet $(\Psi_B)$. Leading derivative couplings proportional to $\partial^\mu\Phi$ are completely determined by chiral symmetry. Next-to-leading order terms come with additional low-energy constants that need to be fixed by comparison with experiment.  Explicit chiral symmetry breaking is incorporated through meson 
(or quark) mass terms.

The chiral SU(3) coupled-channels approach has been applied successfully to describe threshold $\bar{K}N$ physics including the $K^-p$ scattering length and branching ratios. It explains the $\Lambda(1405)$ is an  I = 0 quasibound state emerging from the coupling between the $\bar{K}N$ and $\pi\Sigma$ channels. The chiral interactions that enter the coupled channel matrix are quite strongly attractive in both the diagonal $\bar{K}N$ and $\pi\Sigma$ channels. The non-diagonal $\bar{K}N\leftrightarrow\pi\Sigma$ couplings are also strong and must be treated accordingly. 

{\it The two-poles scenario}.
In the absence of channel couplings, the attraction from the TW matrix elements acting separately in the $\bar{K}N$ channels is sufficient to produce a $\bar{K}N$ bound state with isospin $I=0$ in a narrow window just below $K^-p$ threshold ($\sqrt{s_{thr}} = 1432$ MeV) and above $\sqrt{s} = 1420$ MeV. At the same time, the attraction in the I = 0 $\pi\Sigma$ channel is sufficient to produce an s-wave resonance around $\sqrt{s} \simeq 1390$ MeV, i.e. about 60 MeV above the $\pi\Sigma$ threshold, with a large width of about 200 MeV.  (Note that this resonance is not to be confused with the p-wave $\Sigma(1385)$, the strangeness S = -1 analogue of the $\Delta$ resonance). With the $\bar{K}N \leftrightarrow \pi\Sigma$ channel coupling turned on the $\bar{K}N$ quasibound state moves to $\sqrt{s} \simeq 1420$ MeV and develops a width of about 40 MeV for the decay into $\pi\Sigma$. The $\pi\Sigma$ resonance itself moves slightly upward in energy and reduces its width to 100-150 MeV. 
 
\begin{figure}[htb]
\begin{minipage}[t]{65mm}
\includegraphics[width=6.4cm]{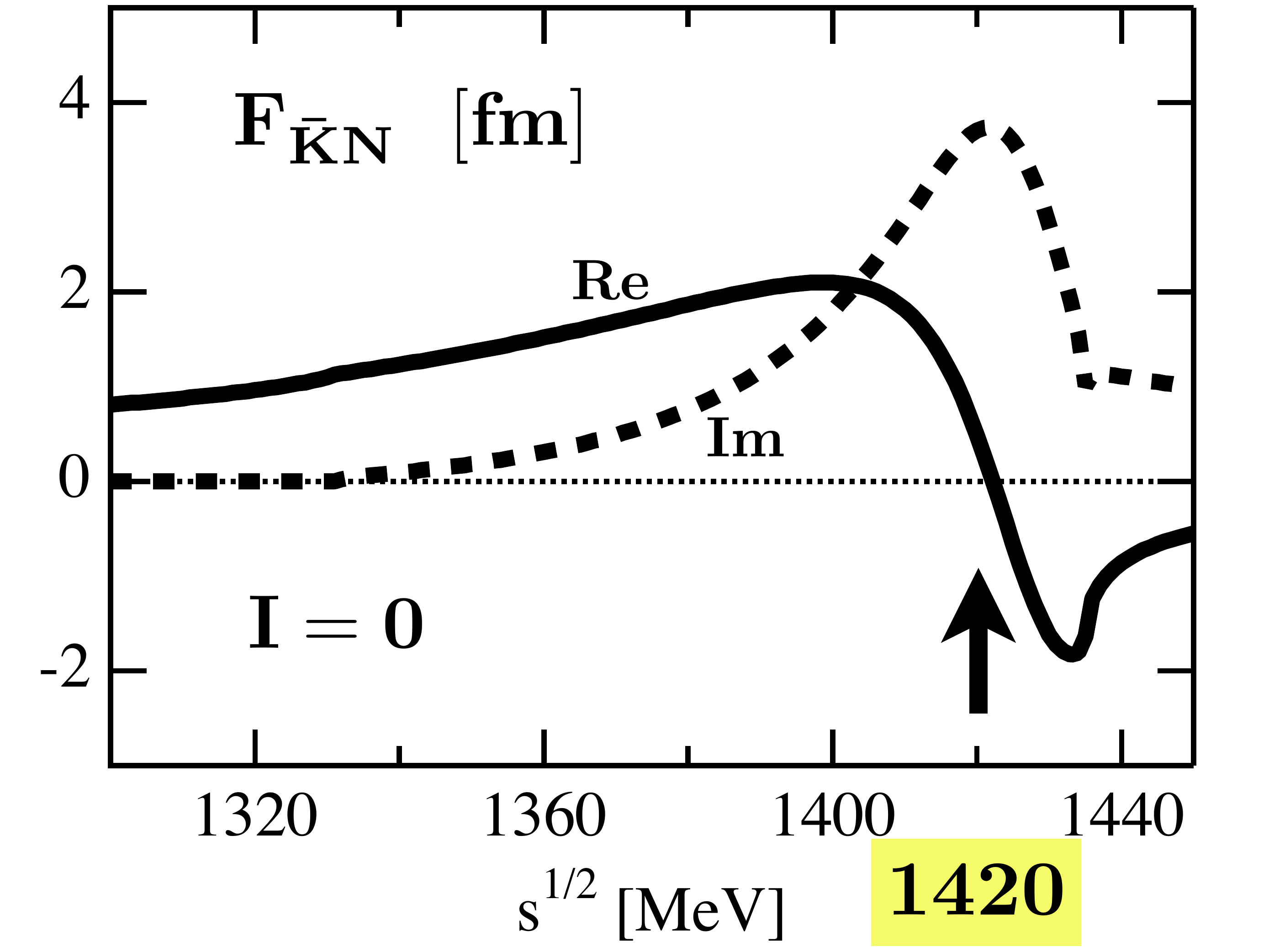}
\end{minipage}
\hspace{\fill}
\begin{minipage}[t]{65mm}
\includegraphics[width=6.5cm]{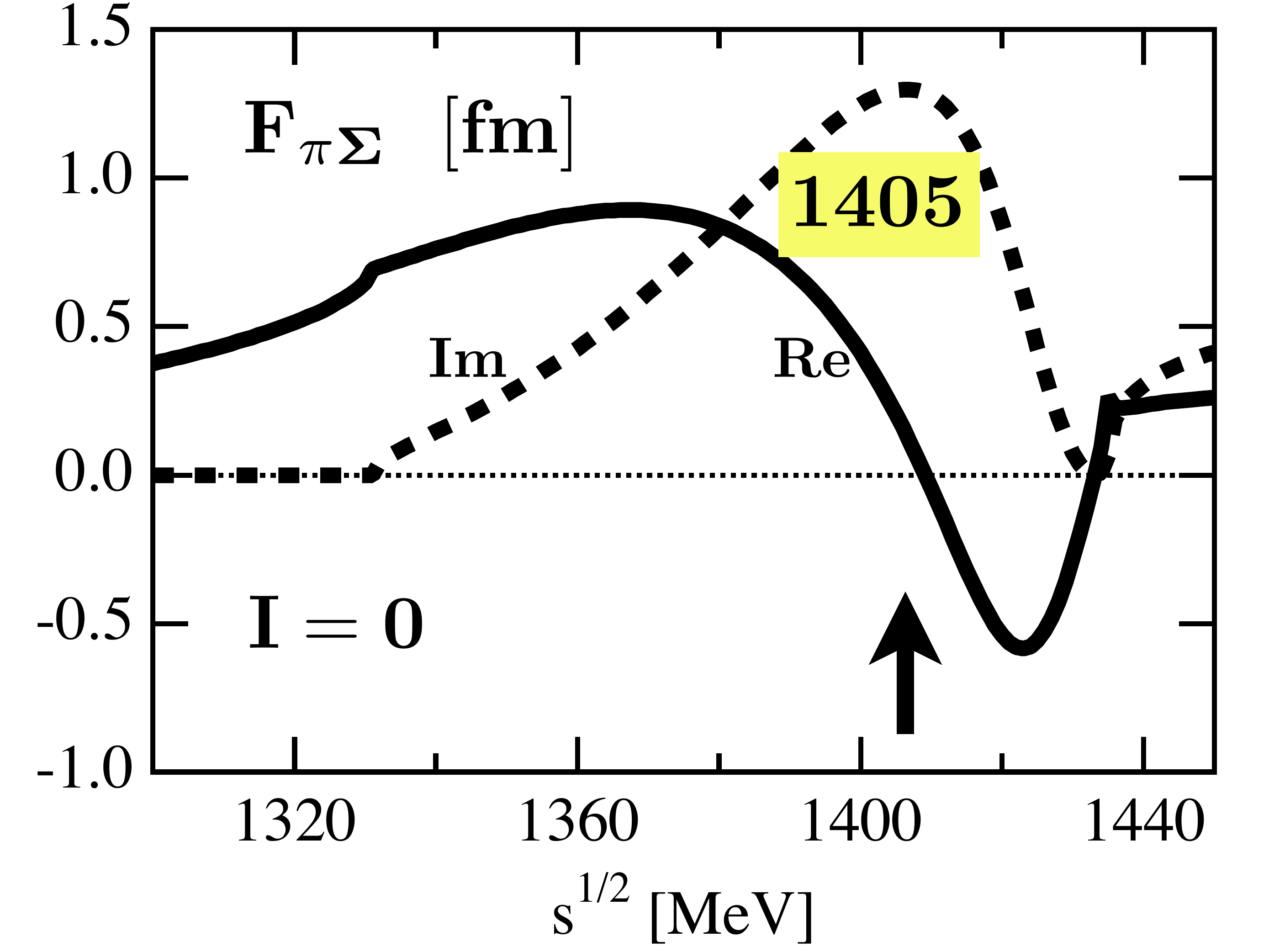}
\end{minipage}
\caption{Real and imaginary parts of the I = 0 (off-shell) $\bar{K}N$ (left) and $\pi\Sigma$ (right) forward amplitudes calculated using the chiral SU(3) coupled-channels approach \cite{HW08}.}
\label{fig:2}
\end{figure}

 This two-poles scenario \cite{Jido03} implies that the $\Lambda(1405)$ is not a simple bound state but must be seen as emerging from the complex interplay of these strongly coupled channels. While the maximum of the $\pi\Sigma$ invariant mass spectrum, commonly identified with the $\Lambda(1405)$, is indeed located around 1405 MeV, this does not mean that the same maximum is to be found in the imaginary part of the (not directly observable) $\bar{K}N$ subthreshold amplitude. This is evident from a recent analysis \cite{HW08}, results of which are reproduced in Fig.\ref{fig:2}. While the $\pi\Sigma$ amplitude has its (observable) imaginary part peaking at 1405 MeV, the $\bar{K}N$ amplitude displays its pole position at 1420 MeV.  

{\it Subthreshold antikaon-nucleon effective interaction}. When ``integrating out" the $\pi\Sigma$ degrees of freedom, this coupled-channels scenario is the basis for the construction of an effective, energy dependent, non-local  subthreshold $\bar{K}N$ interaction \cite{HW08}. This interaction has little in common with the strongly attractive, local and energy-independent Akaishi-Yamazaki (AY) potential Ref.\cite{AY02} used in their calculations of deeply bound $\bar{K}$-nuclear clusters. Although the amplitudes derived from both interactions roughly agree, by construction, in the vicinity of the $\bar{K}N$ threshold, the effective interaction deduced from chiral SU(3) dynamics turns out to be far less attractive  than the AY potential in the deep subthreshold region. This just underlines the general uncertainties associated with off-shell extrapolations of the $\bar{K}N$ amplitude.  

\subsection{Constraints and extrapolations}

{\it Threshold $K^-p$ data.} Precision measurements of kaonic hydrogen and their analysis extracting the real and imaginary parts of the  $K^-p$ scattering length set important quantitative constraints for chiral SU(3) dynamics. The best data so far have been obtained at LNF (the DEAR experiment \cite{Beer05}) and earlier at KEK (the PS-E228 experiment \cite{Iwa97}). The  $K^-p$ scattering lengths deduced from these measurements are not fully consistent:
\begin{center}
$a(K^-p) = -0.47\,(\pm\, 0.10) + i\,0.30\, (\pm\,0.17) ~[\mbox{fm}]~~\mbox{[DEAR]}~,$\\
$a(K^-p) = -0.78\,(\pm\, 0.18) + i\,0.49\, (\pm\,0.37) ~[\mbox{fm}]~~\mbox{[KEK]}~.$
\end{center}
It is thus important to resolve this issue at the higher level of precision to be reached with the
SIDDHARTA experiment at LNF \cite{SID06}.

\begin{figure}[htb]
\begin{center}
\includegraphics[width=6.5cm]{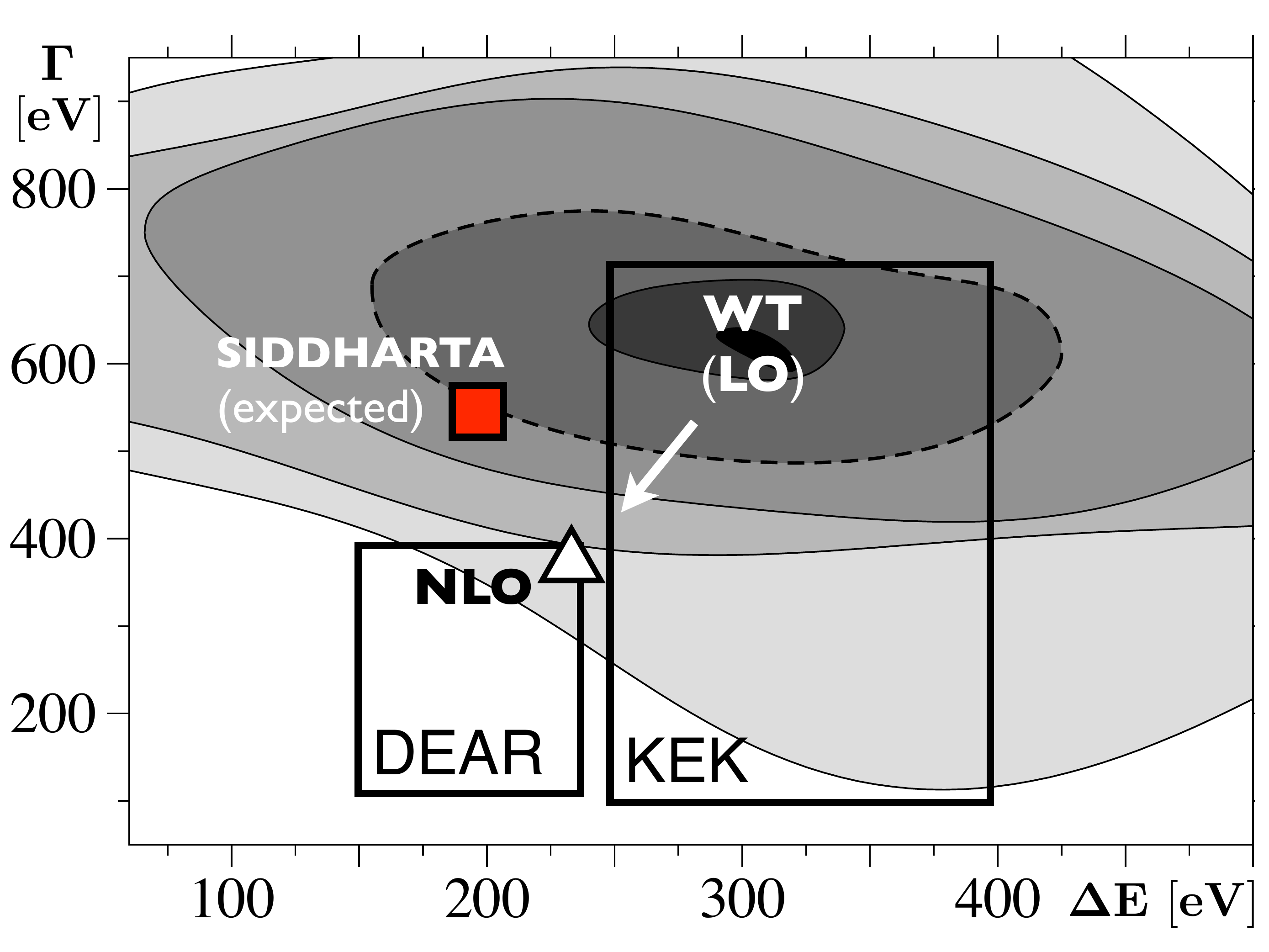}
\end{center}
\caption{Energy shift $\Delta E$ and width $\Gamma$ of kaonic hydrogen. Measured values from DEAR and KEK experiments are shown including errors. Results of chiral SU(3) coupled channels calculations \cite{BNW05,BMN06} using leading order Weinberg-Tomozawa (WT) terms are presented as shaded areas representing different $\chi^2$ upper limits. The $1\sigma$ confidence level is bordered by the dashed line. The next-to-leading order (NLO) result constrained by DEAR data is also shown. The expected precision of the SIDDHARTA experiment is indicated for orientation.}
\label{fig:3}
\end{figure}

Theoretical analyses of the kaonic hydrogen energy shift ($\Delta E$) and width ($\Gamma$), based on chiral SU(3) dynamics, have been performed in Refs.\cite{BNW05,BMN06}. The translation of  $\Delta E$ and $\Gamma$ into real and imaginary parts of the $K^-p$ scattering length now routinely involves corrections \cite{MRR04} beyond the time-honored Deser-Trueman formula. The present situation is summarized in Fig.\ref{fig:3}. The calculations favour slightly the earlier KEK data. With inclusion of higher order corrections in the chiral effective Lagrangian, agreement with the DEAR data can be enforced, but at the expense of loosing overall consistency with $K^- p$ scattering data close to threshold \cite{BNW05}.

{\it Constraints from $\pi\Sigma$ mass spectra}. Extrapolations of the $\bar{K}N \leftrightarrow \pi\Sigma$ coupled-channels dynamics into regions below the $K^- p$ threshold can in principle be tested by examining in detail the shapes and locations of the three $\pi^+\Sigma^-,~\pi^-\Sigma^+$ and $\pi^0\Sigma^0$ invariant mass distributions. It has been pointed out \cite{BNW05} that, enforcing agreement with the DEAR threshold data, these resulting mass distributions would differ from those constrained only by scattering data and threshold branching ratios. This is demonstrated in Fig.4 which emphasizes the need for more accurately determined $\pi\Sigma$ mass spectra. Note also the characteristic difference between $\pi^-\Sigma^+$ and $\pi^+\Sigma^-$ distributions. This splitting is primarily an effect of the $I=1$ component of the amplitude as it interferes with the dominant $I=0$ part \cite{Na99}. The  $\gamma\,p\rightarrow K^+\Lambda(1405)$ photoproduction experiment at JLab \cite{MS09} shows such features although the observed splitting pattern of the $\pi\Sigma$ invariant mass distributions still requires a more detailed understanding.  
\begin{figure}[htb]
\begin{minipage}[t]{65mm}
\includegraphics[width=6.5cm]{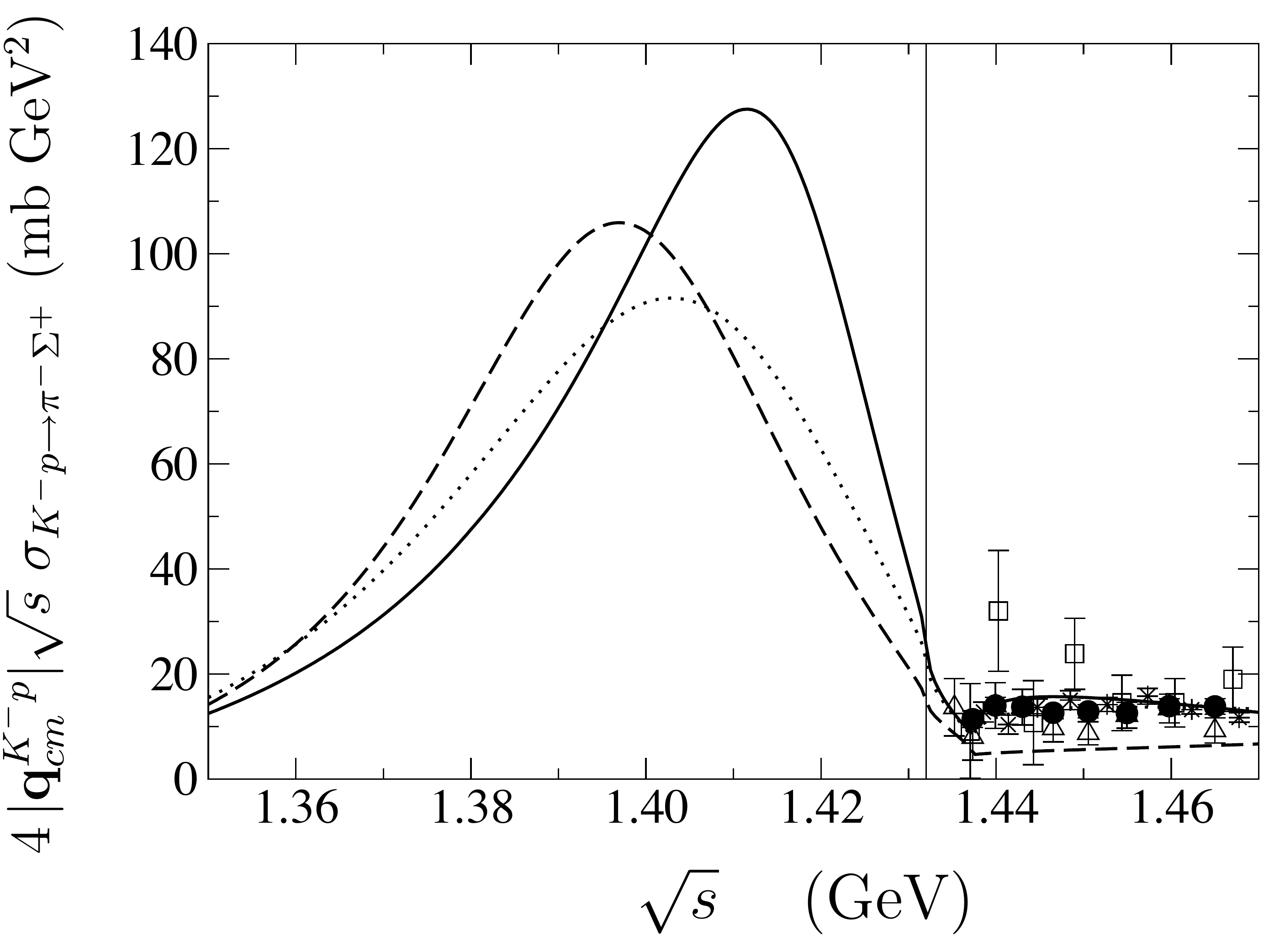}
\end{minipage}
\hspace{\fill}
\begin{minipage}[t]{65mm}
\includegraphics[width=6.5cm]{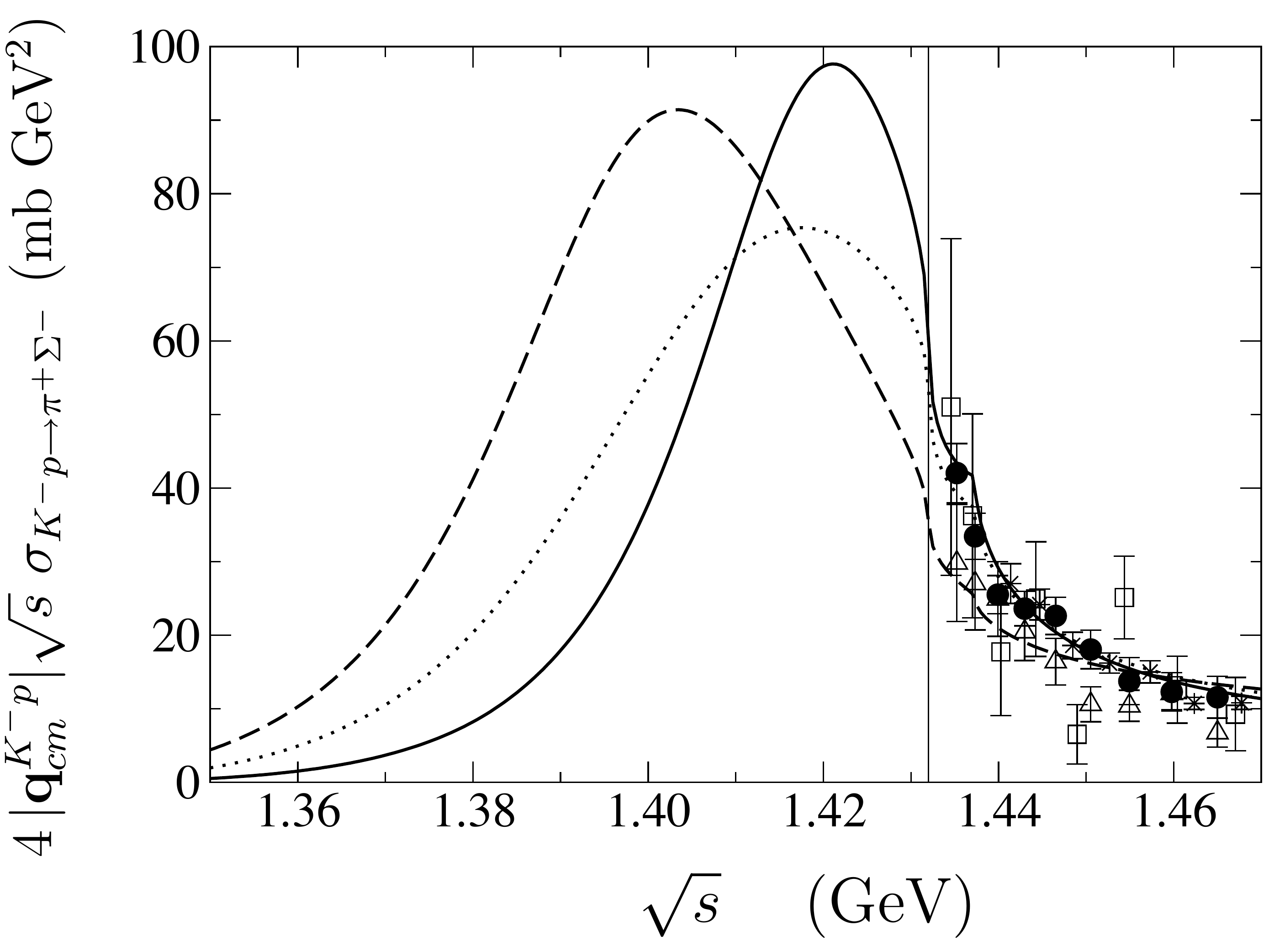}
\end{minipage}
\caption{Subthreshold extrapolations of calculated $K^-p\rightarrow \pi^-\Sigma^+$ (left) and $K^-p\rightarrow \pi^+\Sigma^-$ cross sections \cite{BNW05}. Solid curves: constrained by scattering and threshold observables but without DEAR data; dashed curves: with inclusion of DEAR constraints.}
\label{fig:4}
\end{figure}

\begin{figure}[htb]
\begin{minipage}[t]{65mm}
\includegraphics[width=6.5cm]{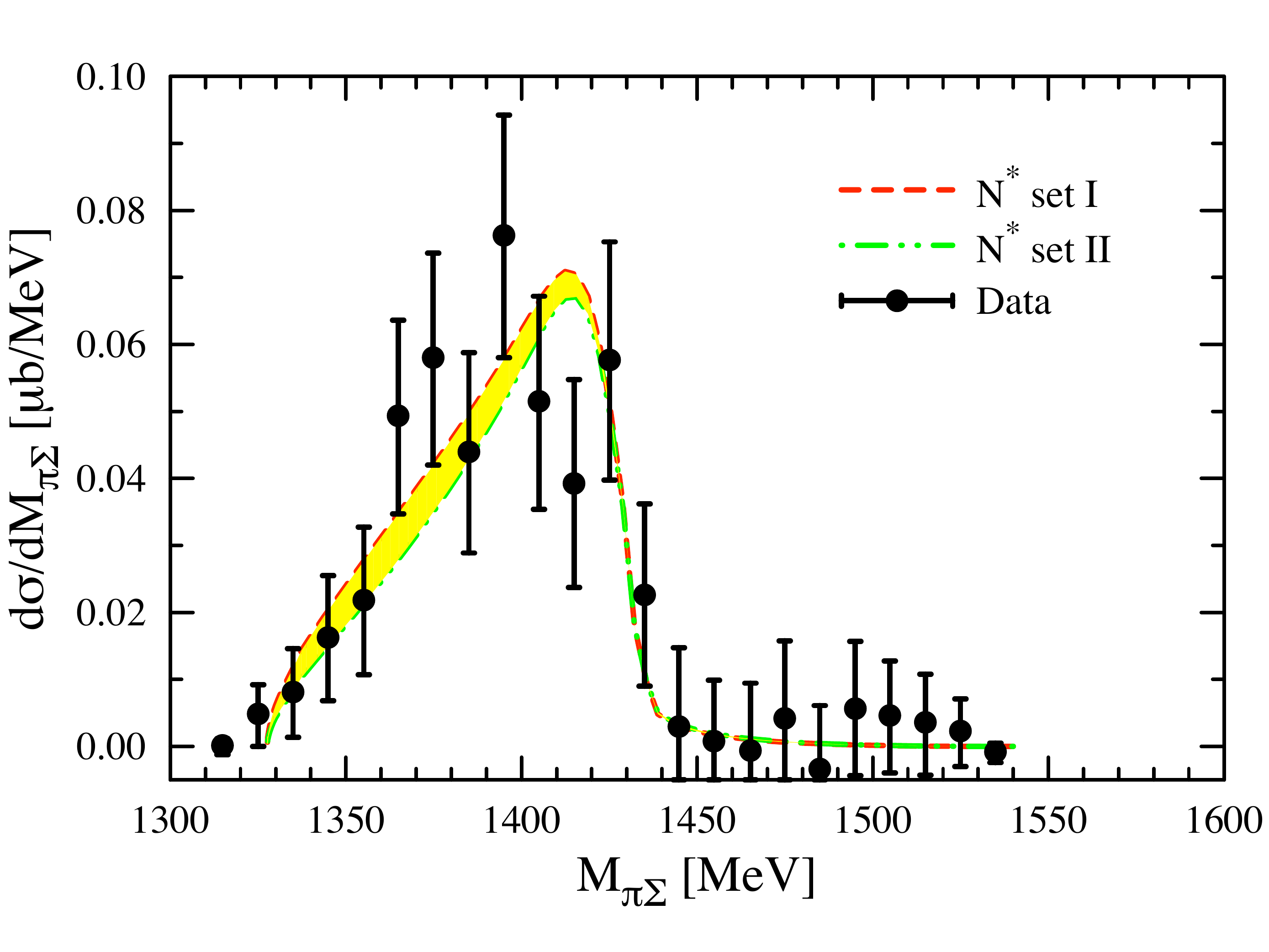}
\caption{The $\pi^0\Sigma^0$ invariant mass spectrum observed in the
$p\,p\rightarrow p\,K^+\,\pi^0\Sigma^0$ reaction (ANKE \cite{ANKE08}) and compared with a calculation \cite{GO07} using chiral SU(3) dynamics.}
\label{fig:5}
\end{minipage}
\hspace{\fill}
\begin{minipage}[t]{6.5cm}
\includegraphics[width=6.3cm]{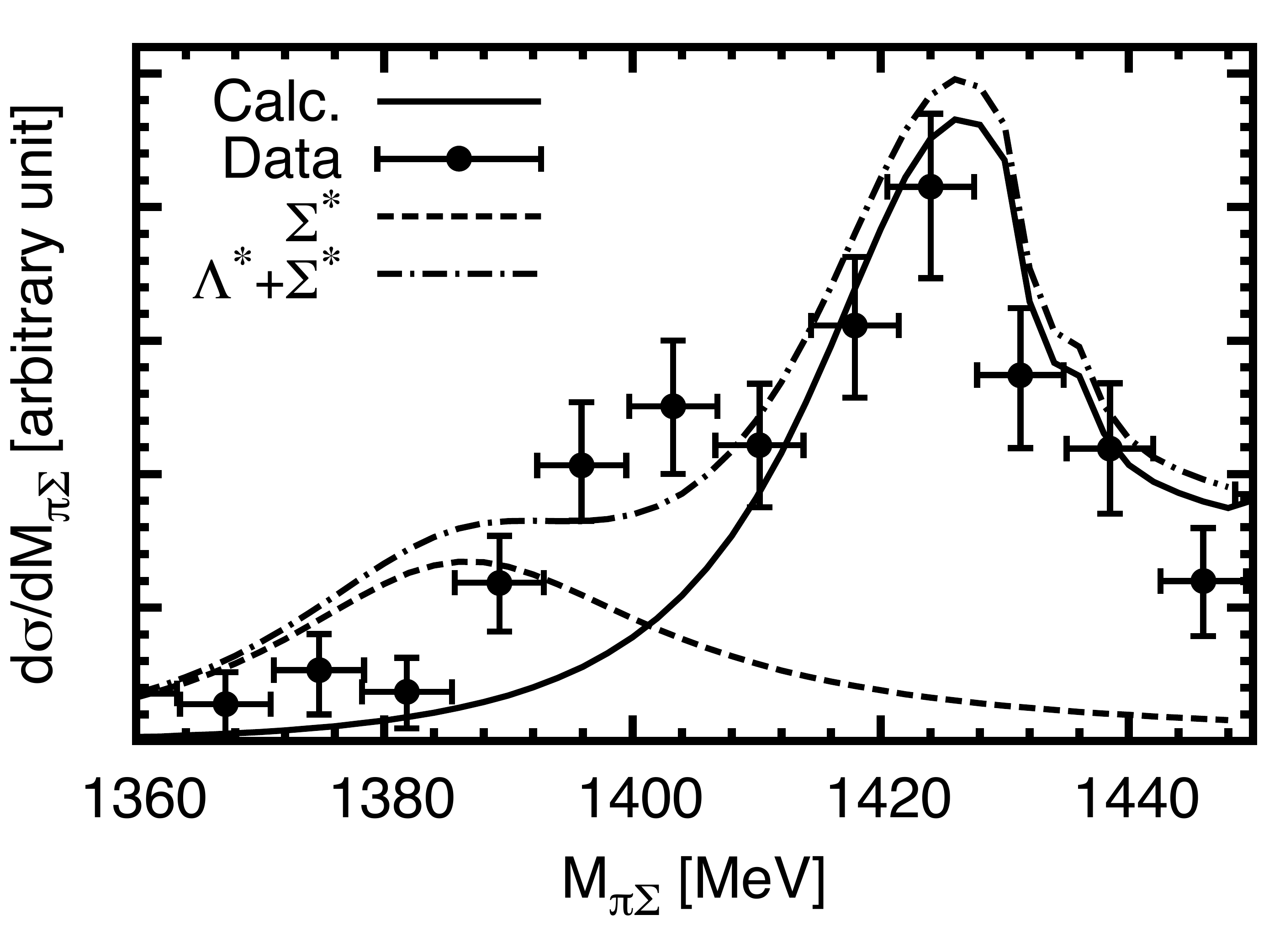}
\caption{The $\pi\Sigma$ invariant mass spectrum deduced from the $K^-d\rightarrow \pi\Sigma\,n$ reaction \cite{Br77} and compared with a calculation involving the chiral $SU(3)$ two-poles scenario \cite{JOS09}.}
\label{fig:6}
\end{minipage}
\end{figure}

A measurement of the $\pi^0\Sigma^0$ mass spectrum by the ANKE experiment, $pp\rightarrow pK^+(\Sigma^0\pi^0)$ \cite{ANKE08} is well reproduced by a calculation based on chiral SU(3) dynamics \cite{GO07} (see Fig.\ref{fig:5}). Likewise, the $\pi\Sigma$ invariant mass distribution observed in $K^-\,d\rightarrow \pi\Sigma\,n$ is well described \cite{JOS09} by the two-poles coupled-channels scenario which, in this case, develops pronounced spectral strength around 1420 MeV rather than 1405 MeV (see Fig.\ref{fig:6}).

It should have become clear from these observations and discussions that there is not just a single $\pi\Sigma$ mass spectrum determining uniqely the position and width of the $\Lambda(1405)$. The $\pi\Sigma$ mass distributions depend on the process considered, a feature quite familiar from reactions involving strongly coupled channels.  

\section{Prototype antikaon-nuclear few-body system: $K^-pp$}

Much recent theoretical activities have been focused on the simplest antikaon-nuclear system, the $\bar{K}NN$ cluster with total isospin $I= {1\over 2}$. The $K^-pp$ system with its $I=1$ nucleon pair, in particular, is a prototype case for studying the role of the antikaon as a possible mediator to bind two baryons which would otherwise not form a bound state.
This is a multichannel three-body problem in which the $K^-pp, \bar{K}^0 pn$ channels are coupled to various charge combinations of $\pi\Sigma N$ and $\pi\Lambda N$. In addition, the $\bar{K}NN \rightarrow YN$ two-body absorption channel must be considered with $Y = \Lambda, \Sigma$. Altogether this is evidently a formidable challenge. A prominent feature is again the $\Lambda(1405)$ appearing in the subthreshold $I=0$ $\bar{K}N$ s-wave. The p-wave interaction involving 
$\pi\Sigma\leftrightarrow\Sigma(1385)$ resonance formation also contributes but has turned out to be less important \cite{DHW09,WG09}.   

Experimental searches for quasibound $\bar{K}NN$ clusters are rewiewed elsewhere in these proceedings. Our primary aim here is to summarize the present status of the theory. Two basic strategies have been employed so far in order to deal with the $K^-pp$ system:
\begin{itemize}
\item{Three-body calculations solving Faddeev equations with separable interactions \cite{SGM07,IS07,IS09};}
\item{ Variational calculations using phenomenlogical input \cite{YA02, WG09} or $\bar{K}N$ effective interactions based on chiral SU(3) dynamics \cite{DHW08,DHW09}.}
\end{itemize}
A survey of computed $K^-pp$ binding energies and widths is given in Table \ref{tab:1}. Even though all input interactions in these calculations have been tuned to reproduce threshold $\bar{K}N$ observables or the $\Lambda(1405)$, one encounters a broad band of binding energies ranging between about 20 and 80 MeV, while the decay widths cover values between 40 and 110 MeV. These results for $\Gamma$ include the $\bar{K}NN \rightarrow \pi\Sigma N$ transitions but not the $\bar{K}NN \rightarrow YN$ decay channels which would further increase the width.

\begin{table}[t]
\begin{center}
\begin{tabular}{lrrrrrrrrrr}
\hline
 &   & Var \cite{YA02} & &  Var \cite{DHW08,DHW09} & &  Var \cite{WG09} &  &  Fad \cite{SGM07} & &  Fad \cite{IS07,IS09}  \\
\hline 
B [MeV] &~~~ & 48  &~~~~~~~~ &   20$\pm$3 &~~~~~~~~  & 40-80  &~~~~~~~~  & 50-70 &~~~~~~~~  & 45-80   \\
$\Gamma$ [MeV] &~~~ & 61 &~~~~~~~~   & 40-70  &~~~~~~~~  & 40-85  &~~~~~~~~  & 90-110 &~~~~~~~~  & 45-75  \\
\hline  
\end{tabular}
\end{center}
\caption{\label{tab:1} 
Binding energies (B) and widths ($\Gamma$) of the quasibound $K^-pp$ system resulting from different variational (Var) and Faddeev (Fad) calculations. The approaches \cite{IS07,IS09,DHW08,DHW09} work with chiral SU(3) based interactions. All other refs. use phenomenological input.}
\end{table}

The wide range of theoretical binding energies reflects on one hand the uncertainties encountered in the subthreshold extrapolations of the $\bar{K}N$ interaction.  On the other hand, both types of calculations, variational or Faddeev, have their intrinsic limitations. Variational calculations use effective $\bar{K}N$ interactions in which the $\pi\Sigma$ channels are eliminated and their effects are relegated to the non-locality and energy dependence of the off-shell $\bar{K}N$ two-body amplitude. But this procedure does not account  for specific features of the full three-body $\bar{K}NN \leftrightarrow \pi YN$
coupled-channels problem. This is handled correctly in the Faddeev approach. As pointed out in Ref.\cite{IS09}, proper treatment of the $\pi\Sigma N$ intermediate state, including the recoiling spectator nucleon as it accompanies the $\pi\Sigma$ subsystem, enhances the binding in the three-body system. This may explain the difference between the variational results of Ref.\cite{DHW08,DHW09} (the ones constrained by chiral SU(3) dynamics) and the Faddeev calculations (of which \cite{IS07,IS09} also use chiral SU(3) input).  In principle the Faddeev method is indeed superior to the variational approach in dealing with the full three-body $\bar{K}NN$ dynamics. On the other hand, solving the three-body equations in practice involves a separable approximation for the basic two-body interactions. While sufficiently many terms are used in the separable expansion of the $NN$ interaction, the driving $\bar{K}N$ interaction is commonly approximated just by a one-term separable form. This may be a source of uncertainty in off-shell extensions of the relevant amplitudes.

As mentioned, additional broadening of the $K^-pp$ quasibound state is expected to come from $\bar{K}$ absorption procsses on two nucleons leading to hyperon-nucleon pairs. Progress has recently been made \cite{SJK09} in obtaining an improved estimate for this non-mesonic width in nuclear matter.
The reaction considered is $K^-NN \rightarrow \Lambda(1405)\,N \rightarrow Y N$. Including pion, kaon and eta exchange mechanisms in their calculations \cite{SJK09}, the authors find a width $\delta\Gamma(\Lambda^* N\rightarrow Y N) \simeq 22$ MeV at normal nuclear matter density. For the $K^-pp$ quasibound state with its lower density,  a simpler estimate \cite{DHW09} points to a corresponding added width of $\delta\Gamma \sim 10$ MeV. In any case, the lifetime of the $K^-pp$ state, if existent,
is expected to be very short.

\section{Heavier antikaon-nuclear systems}

Several exploratory studies have been performed investigating antikaon-nuclear binding in heavier nuclei. One such approach solves the Klein-Gordan equation with a $\bar{K}$-nuclear complex potential based on chiral SU(3) dynamics and including Pauli plus short-range NN correlations \cite{WH08}. An alternative approach uses relativistic mean field theory with nucleons, scalar-vector mean fields and kaons as a framework for self-consistent calculations of antikaonic nuclei with one or more $K^-$ bound to the nuclear core \cite{MFG06,GFGM07}. Binding energies in such systems can reach 100 MeV and beyond, but always accompanied by large widths, larger than 50 MeV, once $K^-$ absorption on two nucleons is taken into account \cite{WH08}. An interesting question concerns possible enhancements of binding effects for multi-antikaonic nuclei. The investigations of Ref.\cite{GFGM08} with inclusion of hyperons show however that there is saturation with increasing kaon number, making kaon condensation in dense matter unlikely.

\section{Conclusions}

High-precision $\bar{K}N$ threshold data and accurate $\pi\Sigma$ mass spectra are crucial in order to constrain subthreshold extrapolations of  the antikaon-nucleon interaction into domains relevant for possible $\bar{K}$-nuclear quasibound states. Establishing improved $I=0$ and $I=1$  $\bar{K}N$ scattering lengths is a step of prime importance in this context.

A necessary (though not sufficient) condition for reliable subthreshold extrapolations is a controlled theoretical framework. Chiral SU(3) effective field theory combined with coupled-channel methods provides such a framework, but it requires a sufficiently large and accurate empirical data base in order to proceed.

While unambiguous conclusions about quasibound antikaon-nuclear systems can at present not yet be drawn, further progress is expected to come from detailed investigations of exclusive final states following $K^-$ absorption and photon- or hadron-induced $K^+$ production on nuclei, in order to constrain the underlying coupled-channel dynamics.\\

\newpage

{\it Acknowledgements}  \\

Sincere thanks go to my former collaborators on topics related to this report, Akinobu Dot\'e and Tetsuo Hyodo. Stimulating discussions with Avraham Gal and his careful reading of the manuscript are gratefully acknowledged.

\end{document}